%% file: main.tex
\documentclass[11pt,a4paper]{article}

\usepackage[utf8]{inputenc}
\usepackage[T1]{fontenc}
\usepackage{lmodern}
\usepackage[margin=2.4cm]{geometry}
\usepackage[english]{babel}
\usepackage{amsmath,amssymb,amsthm}
\usepackage{booktabs,tabularx,array}
\usepackage[dvipsnames]{xcolor}
\usepackage{graphicx}
\usepackage{tikz}
\usepackage{pgfplots}
\usepgfplotslibrary{groupplots}
\pgfplotsset{compat=1.18}
\usepackage[round,authoryear,sort&compress]{natbib}
\usepackage[colorlinks=true,linkcolor=MidnightBlue,citecolor=MidnightBlue,urlcolor=MidnightBlue,hypertexnames=false]{hyperref}
\hypersetup{
  pdftitle={Literary Axioms: A Conceptual Framework for Literary Creation, Interpretation, and Evaluation},
  pdfauthor={Qiang Liu; Chunyi Zhao},
  pdfsubject={A conceptual theory of literary configurations, realization, and reader reconstruction},
  pdfkeywords={literary theory, literary creation, interpretation, literary evaluation, computational creativity}
}
\usepackage{enumitem}
\usepackage{float}
\usepackage{microtype}
\usepackage{xspace}

\newtheorem{definition}{Definition}
\newcommand{\system}{\textsc{Literary Axioms}\xspace}
\newcommand{\axiom}[1]{\texttt{ax\_#1}}
\newcolumntype{Y}{>{\raggedright\arraybackslash}X}

\title{\textbf{Literary Axioms: A Conceptual Framework\\for Literary Creation, Interpretation, and Evaluation}}
\author{
  Qiang Liu\thanks{Corresponding author: \texttt{will.liuqiang@gmail.com}}\\
  \textit{Noevara Inc.}
  \and
  Chunyi Zhao\\
  \textit{Centre of Educational Design and Learning, University of Otago}\\
  \texttt{cccchunyi07@gmail.com}
}
\date{September 2026}

\begin{document}
\maketitle

\begin{abstract}
How can the conceptual commitments of a literary work connect its creation,
interpretation, and evaluation? We propose Literary Axioms, a conceptual framework
in which a work develops a configuration of claims about experience and commitments
to literary form. A literary axiom is a provisional organizing premise whose scope
and textual realization can be examined. The framework develops Qiang Liu's
proposal that writers select and combine such premises, realize them through
characters, events, language, and form, and make them available for reconstruction
by readers. We distinguish an open axiom space, a selected configuration, its
relational organization, and situated reader reconstructions. This yields an account
of coherence as organized compatibility or conflict, originality as a change in
claims, relations, or realization, and interpretive richness as a plurality of
textually grounded reconstructions. Five dimensions connect this account to
literary evaluation: coherence, contextual reach, distinctiveness, realization
fidelity, and interpretive richness. We specify boundary cases and derive five
empirical predictions with observations that would count against them. A concrete
implementation contains 1,455 Chinese-language records, 1,464 mappings to 149 works,
and 472 typed relationships. A descriptive audit establishes the artifact's
structural integrity and identifies gaps in reuse, context coding, and provenance.
The paper contributes a theory-led framework, an inspectable implementation, and a
research program; its predictions await reader and creative-writing studies.
\end{abstract}


\section{Introduction}
\label{sec:introduction}

A writer's decisions about characters, conflict, voice, and endings often develop
a conception of human experience. A reader works in the other direction, using
those decisions to reconstruct what a work makes thinkable. Literary evaluation
then asks how convincingly and distinctively the work sustains these possibilities.
This paper proposes a common conceptual account of these three activities.

The starting point is Qiang Liu's literary-axiom proposal: a work can be approached
as the realization of a selected configuration of conceptual and formal premises.
Its original Chinese exposition, \emph{The Mathematical Principles of Literary
Creation: Reflections on Five Years of Literary Reading} (title translated by the
authors), connected a space of potential axioms to the selection of a sub-axiom
system, literary realization, reader reconstruction, and five dimensions of value.
We develop this proposal into an explicit framework with defined objects,
mechanisms, scope conditions, and empirical predictions.

The organizing question is how a configuration becomes literary. A proposition
about desire does not itself specify a character, a scene, or a narrative voice.
Conversely, two stories about the same event can organize incompatible understandings
of it. Our central claim is that the selection of premises, their relations, and
their realization together provide a useful intermediate level between an abstract
theme and the particular text. At that level, a writer can revise an organizing
commitment, a reader can justify a reconstruction, and a critic can compare how
different works embody related concerns.

For example, the claim that desire pursues an idealized past is more specific than
the theme ``desire,'' yet leaves many literary realizations possible. A retrospective
narrator may render that pursuit tragic; a competing voice may expose its costs to
others. Relations among claims and choices of form thus matter as much as membership
in a premise set. The framework makes these dependencies explicit.

The mathematical inspiration lies in organizing premises and examining their
consequences. Literary axioms remain provisional, situated commitments expressed in
natural language. This use supports conceptual analysis without requiring a proof
calculus. It also accommodates works that question their own premises and readers
who recover meanings beyond a writer's plan.

The paper makes four contributions:
\begin{enumerate}[leftmargin=2em]
  \item Four connected theoretical claims about configuration, realization,
  reconstruction, and comparison, grounded in the original literary-axiom proposal.
  \item A model of selection, relational organization, textual realization, and
  situated reading, with five evaluative dimensions and explicit boundary cases.
  \item A concrete representation and descriptive audit of 1,455 records, together
  with work-level skeletons that illustrate the model's implementation.
  \item Five testable predictions and a staged validation program that separates
  record quality from evidence about interpretation and creative practice.
\end{enumerate}

Sections~\ref{sec:claims}--\ref{sec:boundaries} develop the claims, mechanisms, and
limits. Sections~\ref{sec:dataset}--\ref{sec:mapping} instantiate them in a dataset
and work-level representations. Section~\ref{sec:predictions} specifies empirical
tests; Section~\ref{sec:related} positions the proposal relative to existing work.

\section{Core Theoretical Claims}
\label{sec:claims}

The following claims specify the explanatory commitments of the framework. They
are conceptual proposals whose practical consequences are tested by the predictions
in Section~\ref{sec:predictions}.

\paragraph{T1: Configurational organization.}
A work's conceptual organization can be investigated through a selected set of
premises and the relations that make them jointly significant. Individual premises
may be inherited; the configuration can still be distinctive. The claim applies to
an identifiable interpretive layer of a work, not to every feature of its language
or experience. It motivates asking what changes when one premise or relation is
removed, qualified, or replaced.

\paragraph{T2: Realization mediates meaning.}
A configuration becomes literary through characters, events, affect, language,
and form. These means do more than illustrate a conclusion: they can qualify,
test, or transform a premise. Coherence and fidelity therefore concern the
relationship between the configuration and its realization. A character's assertion,
a narrator's assertion, and a commitment made available by the work require separate
interpretive judgments.

\paragraph{T3: Reading reconstructs and differentiates.}
Readers reconstruct configurations through textual evidence and their own
experience, literary knowledge, and historical situation. Reconstruction can recover
an organizing commitment while altering its scope or relations. The creation--reading
cycle thus links two activities without making reading a recovery of a single
authorial code. Grounded differences between reconstructions constitute one source
of interpretive richness.

\paragraph{T4: Evaluation is relational and multidimensional.}
Coherence, contextual reach, distinctiveness, realization fidelity, and interpretive
richness identify different achievements and failures. Originality can arise from
a new premise, a new combination or relation, or a new realization of inherited
material. Evaluation requires a stated comparison corpus and reader context;
neither a count of premises nor a single context label determines literary value.

Together, T1 supplies the object of analysis, T2 explains its textual mediation,
T3 connects it to reading, and T4 organizes comparison and evaluation. A useful
application must make a specific interpretive or creative decision clearer than
would a list of themes alone. This requirement provides a practical way to challenge
the framework rather than merely adopt its terminology.

\section{Concepts and Mechanisms}
\label{sec:framework}

\subsection{From an open axiom space to finite repositories}
\label{sec:axiom-space}

The conceptual model distinguishes potential premises from their recorded formulations. Let $\mathcal{U}$ denote an open-ended space of candidate literary
axioms: proposition-like claims about human experience together with formal
strategies through which literature can organize perception, narration, language,
and affect. The original essay denoted this horizon by $S$; we use $\mathcal{U}$ to distinguish it from work-level skeletons.
$\mathcal{U}$ is not assumed
to be enumerable, complete, culture-neutral, or fixed across history. It is a
conceptual horizon. A versioned repository $\mathcal{A}_v$ is instead a finite,
revisable collection of records. Let $q(a)$ extract the candidate commitment
expressed by a record $a$; its represented pool is
$Q_v=\{q(a):a\in\mathcal{A}_v\}\subseteq\mathcal{U}$.

For a creative project $p$, a writer may select a working premise set
\[
  P_p \subseteq Q_v \cup N_p, \qquad N_p\subseteq\mathcal{U}\setminus Q_v,
\]
where $N_p$ contains newly formulated candidates not yet represented in the
repository. The original account called such a configuration a ``sub-axiom
system.'' The present framework retains that idea but does not require the members
of $P_p$ to satisfy formal logical independence. Their relations must
instead be made explicit: some may complement one another, some may be redundant,
and some may be placed in a deliberate tension that the work stages rather than
resolves. Internal coherence therefore concerns the intelligibility of the
configuration, including its organized conflicts, rather than the absence of every
contradiction.

The work is then a \emph{realization} of this configuration through choices of
character, event, setting, voice, imagery, rhythm, genre, and other formal means.
This preserves the original claim that a work can be understood as a concrete
manifestation of a philosophical configuration, while avoiding the stronger claim
that the work is exhausted by that configuration. Literary axioms also need not be
stated directly. A work may make a claim available by staging its consequences,
testing it against a counterexample, distributing it across conflicting voices, or
allowing form to expose the limits of what a character or narrator asserts.

This distinction separates two objects that can otherwise be conflated. $P_p$ is the
\emph{generative premise set}: the claims and formal commitments used during
composition, whether consciously articulated or reconstructed from drafts. The
work-level skeleton $S_w$ in Section~\ref{sec:mapping} is an
\emph{interpretive skeleton}: a curator's or reader's recorded account of
what the completed work makes available. Since $S_w$ stores record--role--note
triples, the comparable commitment set is
$Q_w=\{q(a):(a,r,n)\in S_w\}$. Neither authorial intention nor successful
transmission guarantees $P_p=Q_w$.

\subsection{Relational configurations and realization}
\label{sec:mechanism}

\begin{definition}[Literary axiom]
A literary axiom is a provisional organizing premise about represented experience
or literary form, stated with enough scope to support an interpretive or creative
decision. Its formulation is revisable in light of textual evidence, counter-readings,
or a developing work.
\end{definition}

An affect or style label becomes a premise when its organizing role is articulated.
For example, ``grief'' names a concern; ``repeated domestic actions can sustain grief
without explicit recollection'' proposes a relation between experience and form.
This formulation can guide a scene and can be challenged by a reader who finds that
the repetition instead conveys relief or habit. A raw feeling or genre name need
not be forced into a proposition when that translation adds no insight.

For a project $p$, let its working configuration be
\[
  C_p=(P_p,L_p,F_p,\kappa_p).
\]
Here $P_p$ is the selected premise set; $L_p$ records relations among premises;
$F_p$ records planned means of realization such as voice, sequence, imagery, and
rhythm; and $\kappa_p$ specifies attribution and scope. A relation in $L_p$ names
its endpoints, type, and interpretive rationale. Scope can identify the relevant
speaker, narrative level, historical situation, and extent of a claim. A form
premise in $P_p$ expresses a rationale; a choice in $F_p$ specifies how a draft will
attempt to realize it. Neither a relational configuration nor these plans are
currently stored in full by the dataset schema.

We describe realization and reconstruction as relations,
\[
 \mathsf{Realize}\subseteq\mathcal C\times\mathcal W,
 \qquad
 \mathsf{Read}\subseteq\mathcal W\times\mathcal B\times\mathcal K\times\mathcal C,
\]
where $\mathcal C$ is the space of configurations, $\mathcal W$ the set of works,
$\mathcal B$ readers, and $\mathcal K$ reading contexts. $(C_p,w)\in\mathsf{Realize}$
records an argued realization, not an executable generation rule. A tuple
$(w,b,c,C')\in\mathsf{Read}$ records a reconstruction attributed to reader $b$
in context $c$. A configuration can admit several realizations and a work several
reconstructions, including alternatives by the same reader.

\paragraph{Selection and organization.}
Experience, reading, and reflection supply candidates. The writer selects premises,
states their scope, and organizes their relations. Independence is a working
question about redundancy: does removing a premise lose a relevant distinction?
It is not a theorem about logical derivability. Conflict is evaluated with
attribution intact; competing characters can sustain incompatible commitments
within a coherent dramatic design.

\paragraph{Realization and revision.}
To realize a configuration, a writer chooses situations in which commitments have
consequences. These consequences may be actions, exclusions, silences, affective
patterns, or formal effects. Drafting can reveal that an apparent complement is a
tension, prompting revision of $L_p$ or $P_p$. Form can also introduce an implication
not present in the plan. Revision therefore operates on both configuration and text.

\paragraph{Evidence and comparison.}
A reader supports a reconstruction with textual or formal evidence and explains the
inferential step. Comparing configurations requires aligning their meanings,
attributions, and scopes, not merely matching record identifiers. Different wording
can express a similar commitment; identical wording can conceal different scopes.
Realization fidelity asks whether the intended exploration is available in the
text. It can include the deliberate refutation of a character's premise if that
refutation belongs to the organizing design.

\subsection{Content, form, and contextual scope}
\label{sec:taxonomy}

Premises can be compared along content/form and contextual-scope dimensions. Abstraction levels provide a further indexing choice in the implementation.

\paragraph{Category.} \emph{Content} records concern represented human experience, society, morality, emotion, relationships, or existence. \emph{Form} records concern narrative structure, language, perspective, tone, or style. The division is an indexing convenience; many interpretations connect form and content.

\paragraph{Context type.} The intended coding asks whether a claim is explicitly tied to temporal and regional contexts: Type A is context-independent in the record; Type B is temporally bounded; Type C is regionally situated; and Type D is both temporally and regionally situated. These labels describe metadata scope, not truth or literary merit. Section~\ref{sec:audit} shows that the current field population does not reliably distinguish Types C and D, so this part of the taxonomy should be treated as provisional.

The originating essay associated fewer temporal or regional restrictions with
greater potential reach. This motivates a hypothesis about transfer across
contexts, not a demonstrated ranking of works. B and C are not naturally ordered:
a temporal restriction and a regional restriction constrain different axes.
Actual reach requires evidence from readers and reception, rather than inference
from an A--D label.

\paragraph{Abstraction.} Records are labeled \emph{fundamental}, \emph{intermediate}, or \emph{specific}. The labels support filtering at different granularities. Their boundaries are interpretive and have not yet been tested for annotator agreement.

\subsection{Creation and reading as an interpretive cycle}
\label{sec:cycle}

The framework uses the following diagram to connect the relations just defined:
\[
C_p \xrightarrow{\text{composition and realization}} w
\xrightarrow{\text{situated interpretation}} C'_{w,b,c}.
\]
Here $C_p$ is a generative configuration, $w$ is a resulting work, and
$C'_{w,b,c}=(P',L',F',\kappa')$ is one configuration reconstructed by reader $b$
in context $c$. In reconstruction, $F'$ describes perceived formal organization
rather than a writing plan. The premise sets $P_p,P'\subseteq\mathcal U$ may
overlap while their relations, realization accounts, and attributions differ.

The original account called the outward movement ``deduction'' and reader
reconstruction ``induction.'' We retain that directional analogy: composition
develops consequences of selected premises, while reading infers organizing
commitments from their realization. Composition also includes invention,
association, revision, and tacit knowledge; neither arrow specifies a formal
inference calculus. The notation exposes two practical questions: whether selected
premises guide coherent creative choices, and whether readers recover related
commitments from the completed text.

As a creator's heuristic, the cycle can be expanded into four operations. First,
experience, reflection, reading, and research supply candidate claims. Second, the
writer selects and relates a working configuration, checking unintended
contradiction and unnecessary redundancy while preserving productive tension.
Third, the configuration is realized through the work's represented world and
formal organization rather than merely announced as a thesis. Fourth, readers
reconstruct one or more grounded configurations from the completed text. Repeated
movement through these operations can support revision: a writer may alter the
configuration when the draft produces different implications, or alter the draft
when its formal choices do not make the intended questions available.

The combinatorial language of the original theory is retained at this heuristic
level. A writer can create a new configuration by selecting, relating, and
transforming claims that have appeared elsewhere. Originality, however, is not
established by set inequality alone. It is relational to a declared comparison
corpus and may arise from at least four sources: a newly formulated claim, an
unusual combination of inherited claims, a new relation among them, or a distinct
formal realization. Comparing proposed configurations with skeletons reconstructed
from prior works can therefore aid creative exploration, but the comparison must
include context, provenance, and form before it can support a novelty judgment.

The framework also motivates five candidate dimensions for discussion: organized coherence within a configuration, contextual reach, distinctiveness of premises, relations, attribution, or formal realization, fidelity between a working configuration and its textual realization, and richness of grounded reconstructions. These dimensions are prompts for analysis. In particular, broad contextual scope is not a proxy for quality: situated works may be aesthetically and historically important precisely because of their specificity.

The original account also used ``induction'' for recognition of lived experience
in a fictional situation, as well as reconstruction of its underlying claims.
These are distinct reader responses: resonance does not demonstrate a claim's
truth. Reader experience and literary familiarity may affect both responses.
Complex or indirect realization may demand more interpretive work and reach a
smaller audience without thereby being of lower literary value. Reconstructing
prior works likewise does not establish that their authors consciously composed
from an explicit axiom configuration.

\subsection{A worked conceptual example}
\label{sec:concept-example}

The following invented miniature illustrates T1--T3; it is not an observed writing
experiment. Select two candidate premises: $p_1$, ``care can require personal
sacrifice,'' and $p_2$, ``care obligations do not extinguish personal autonomy.'' Place them in
tension, attribute them to different perspectives, and set the scene around an
adult daughter deciding whether to remain at her father's home. Use an object and
an incomplete exchange rather than stating either premise in dialogue:

\begin{quote}
She put the train ticket beneath his medicine box. He slid it back across the
table. ``You will miss it.'' She filled tomorrow's compartments before answering.
Outside, the last bus stopped, waited, and went on.
\end{quote}

The care routine and the ticket make two commitments available through action.
Her delay can suggest sacrifice; his gesture can suggest recognition of her
autonomy. A reader might instead read the gesture as concealed reproach. That
alternative requires an argument about the exchange, not simply a new label.
The miniature leaves the relation unresolved, so its coherence rests on making
the conflict intelligible.

Now keep the broad situation and premise set but change the realization:
\begin{quote}
He watched her put the ticket beneath his medicine box. ``You will miss it,'' he
said, using the voice that had always made her stay. She picked up the ticket.
This time, he could not find the words to stop her.
\end{quote}

The added attribution makes the invitation to leave a means of control. The
daughter's decision to reclaim the ticket changes the relative force of care and autonomy. The premise
inventory alone cannot capture this difference: attribution, relation, and voice
do explanatory work. Replacing the controlling clause with neutral description
would again change the evidence available to a reader.

The example also connects originality to comparison. Neither care nor autonomy is
an unprecedented concern; a distinctive configuration may arise through their
relation, allocation to voices, or realization. Whether these particular passages
are original or effective remains a comparative literary judgment. The example
demonstrates what the representation asks a writer or reader to inspect.

\subsection{Five dimensions of literary evaluation}
\label{sec:dimensions}

The five dimensions originated as a creator's heuristic for asking why a work feels durable or distinctive. They are retained because they generate testable questions, but the framework does not combine them into a scalar ``literary quality'' measure. Each dimension can be desirable in one reading and deliberately violated in another. The following operational questions make the distinction explicit.

The originating account proposed that coherent and independent premises support
credibility, less situated premises support wider resonance, distinctive
configurations support originality, faithful realization connects text to its
conceptual core, and different readers' reconstructions reveal depth and
multiplicity. Here these are research hypotheses, not necessary conditions for
good literature. Logical independence is replaced by a check for unnecessary
redundancy, and coherence is qualified to accommodate deliberate conflict.

\begin{enumerate}[leftmargin=2em]
  \item \textbf{Interpretive coherence.} Can the selected claims be held together after their scope, speaker, and level of abstraction are stated? A work may intentionally stage incompatible ideologies; in that case, coherence may reside in the orchestration of the conflict rather than in agreement among claims. A rating should therefore distinguish an accidental contradiction from a marked dramatic contradiction.
  \item \textbf{Contextual reach.} Does a claim travel beyond its originating historical or regional setting, and what is lost when it travels? A context-independent label is not evidence of universality. The relevant observation is whether readers in a declared context can articulate a meaningful connection, including a connection grounded in difference rather than identification.
  \item \textbf{Distinctiveness.} Does a configuration add a non-redundant perspective through its premises, relations, attribution, or formal realization relative to a declared comparison corpus? Distinctiveness is relational and time-dependent. A configuration can be original to a corpus while drawing on an older philosophical or formal tradition. The dataset currently lacks the comparison corpus and attribution needed for a novelty estimate.
  \item \textbf{Realization fidelity.} Do textual choices repeatedly make the selected claims available, or does the summary merely describe the author's topic? Fidelity concerns the relation between proposition and realization, not whether a critic agrees with the proposition. A claim can be contradicted by a work in a productive way; such a case should be recorded as a counter-reading rather than silently marked as a failed mapping.
  \item \textbf{Interpretive richness.} Can competent readers construct more than one grounded reconstruction without treating every interpretation as equally plausible? Richness is not ambiguity for its own sake. It depends on textual affordances, formal organization, historical knowledge, and the ability to specify where readings converge and diverge.
\end{enumerate}

These dimensions also expose a design tension. A highly explicit record is easier to classify but may flatten the ambiguity that supports richness. A highly abstract record travels across works but may become indistinguishable from a theme. The schema therefore stores statement, description, example, and contextual fields separately so that a concise index can coexist with a longer qualification. Future evaluation should measure this trade-off rather than optimize one dimension in isolation.

The motivating account treated language, style, plot, and technique primarily as
means by which a selected configuration becomes perceptible. The operational model
preserves their realization role but does not rank them as secondary decoration.
Form can generate, qualify, or undo a proposition, and two works that engage a
similar claim set may differ fundamentally in what their formal organization makes
thinkable. This is why form records and technique mappings remain part of the axiom
repository rather than being placed outside the framework.

\section{Scope Conditions and Counterexamples}
\label{sec:boundaries}

The framework applies where a writer or reader can articulate organizing
commitments and connect them to particular textual or formal choices. The work need
not have been composed from an explicit plan. For historical works, a reconstructed
configuration is an interpretive hypothesis unless drafts or other evidence support
an account of the author's process. For contemporary projects, successive plans and
drafts can document how a configuration changes.

An application should state its unit of analysis (passage, poem, narrative, or
performance), its reading context, and the evidence available for its claims.
Where sonic, bodily, or visual experience resists propositional description, the
appropriate output may be a partial configuration with an explicit remainder, or
no useful configuration at all. The framework succeeds through the distinctions
it helps explain, not through the number of records it can assign.

\subsection{Counterexamples that constrain the framework}
\label{sec:counterexamples}

The dimensions are most useful when tested against cases that resist them. A work can be deliberately self-contradictory: a narrator may affirm a moral claim while the form exposes its failure. A context-bound work can achieve wide significance precisely through its historical specificity, so contextual reach cannot be a monotonic quality criterion. A formally innovative work may use a recognizable inherited genre; distinctiveness need not mean that every component is unprecedented. A work can sustain an unresolved ending in which no single axiom set is a faithful summary, making realization fidelity a question about tensions rather than one-to-one correspondence. Finally, a short lyric or highly compressed parable may invite rich readings without supporting a long list of atomic records.

These counterexamples imply three safeguards. Records should retain the scope and evidential note that qualify a proposition; mapping roles should permit technique and content to be separated without pretending they are independent; and disagreement should be represented as data rather than averaged away. In particular, a future benchmark should include deliberately difficult records, negative mappings, and works selected for formal resistance to paraphrase. Otherwise, a high agreement score could simply show that the sample contains easy, culturally familiar claims.

\subsection{Limits of the mathematical metaphor}

Mathematical axioms are premises inside systems whose syntax and inference rules are explicitly defined. Literary axioms in this project are natural-language interpretations. They can be vague, culturally situated, mutually tensioned, or rejected after a new reading. The relationship-strength field is not a truth value, and the composition--interpretation cycle supplies no formal inference procedure. Calling the records axioms is productive only if this distinction remains explicit.

Similarly, the apparent $2^{1455}$ power set of record combinations does not model a genuine literary search space. Most combinations lack semantic coherence; literary form cannot be inferred from set membership; and the current records are neither independent nor complete. The combinatorial analogy may motivate interface design, but it should not be treated as an explanation of creativity.

\subsection{Conditions under which the account would need revision}

The configurational proposal loses explanatory value if equally concise theme
lists support the same distinctions and decisions, or if independent readers
cannot ground the proposed premises in textual evidence. The relational mechanism
is weakened if removing or changing the claimed relations does not affect
interpretation in appropriately controlled cases. The evaluation program must be
revised if its dimensions cannot be distinguished reliably or fail to account for
the judgments they were introduced to illuminate. These are substantive challenges
to particular parts of the account; no single result would settle the value of
literature as a whole.

\section{Data Implementation and Descriptive Audit}
\label{sec:dataset}

\subsection{Record schema}

\begin{definition}[Literary axiom record]
A literary axiom record encodes a candidate premise as a structured record
\[
a=(id,s,d,c,t,l,g,h,e,u,k,E),
\]
where $id$ is a unique identifier; $s$ is a concise proposition; $d$ is an explanatory description; $c$ is a content/form category; $t$ is a contextual-scope type; $l$ is an abstraction level; $(g,h)$ is a domain/sub-domain pair; $e$ and $u$ are era and region context lists; $k$ is a keyword list; and $E$ is a non-empty list of literary examples. Each example names a work and explains how the work motivates the interpretation.
\end{definition}

This definition follows the fields in the current JSON artifact. The term \emph{atomic} is a construction goal: an entry should express one main claim, but atomicity is a judgment to be tested rather than a property guaranteed by the schema. Likewise, an example records a curator's grounding argument; it does not establish consensus about the work.

A literary axiom differs from four nearby concepts. It is more specific than a one-word theme, because it makes a proposition. It differs from a plot template because it need not specify an event sequence. Its organizing function distinguishes its use here from an aphorism presented for standalone effect; a record adds explicit classification and grounding to that use. It differs from a mathematical axiom because it remains defeasible, situated, and open to competing readings.

\subsection{Construct boundaries and formal commitments}
\label{sec:constructs}

The mathematical notation is useful only when its limits are stated. Let $\mathcal{A}=\mathcal{A}_v$ denote the repository at the analyzed version, and retain $\mathcal{W}$ for the set of works. In the notation below, a stored axiom identifier stands for the record it resolves to. A mapping is not a truth assignment but a typed relation
\[
  M \subseteq \mathcal{W} \times \mathcal{A} \times R \times N,
\]
where $R=\{\textsc{primary},\textsc{secondary},\textsc{technique}\}$ and $N$ is the set of possible natural-language grounding notes. A work's skeleton is therefore a projection of $M$ under a particular curation history. It is neither a complete semantic representation of the work nor a claim that all readers should recover the same set.

Three distinctions prevent the framework from quietly importing mathematical guarantees. First, a formal axiom is a premise inside a specified language and deductive system; in this interpretive dataset, a literary axiom record expresses a reading whose wording, scope, and relevance can be challenged. Second, a theme names a broad concern (for example, alienation), whereas an axiom proposes a relation or mechanism (for example, that a person's social identity can depend on economic usefulness). Third, a motif or symbol is a recurrent textual element, whereas an axiom may connect that element to an interpretation but does not require recurrence. The same record can consequently be useful for close reading and for creative planning while remaining inadequate as a substitute for textual evidence.

The term ``self-consistency'' also needs a local definition. A set of claims is consistent in this project when a reader can hold them together as a plausible interpretation of a work without an unmarked change of scope. It does not mean that the claims are true in the world or that a work's characters endorse them. Similarly, a \textsc{contradicts} link means that two claims are difficult to maintain in one declared interpretive frame. It is not a formal proof of $p\land\neg p$. A \textsc{tensions} link is weaker: the claims may coexist as competing perspectives, unresolved conflicts, or historically distinct readings.

Finally, composition and interpretation are intentionally non-inverse operations. A writer can select a claim set and produce a work that complicates, ironizes, or refutes the selected claims. A reader can infer claims that were not part of the writer's initial plan. The cycle in Section~\ref{sec:cycle} therefore describes an interface between activities, not a reversible encoding scheme.

\begin{table}[t]
\centering
\caption{Construct boundaries used in the framework.}
\label{tab:construct-boundaries}
\small
\begin{tabularx}{\linewidth}{@{}lY Y@{}}
\toprule
\textbf{Construct} & \textbf{Operational role} & \textbf{What it does not claim} \\
\midrule
Formal axiom & Premise in a specified formal system & Empirical truth outside the stipulated system \\
Literary axiom & Provisional organizing premise in creation or interpretation & A universal law or complete interpretation \\
Axiom record & Stored formulation, classification, and grounding & Exhaustive realization of the conceptual model \\
Theme & Broad concern used for navigation & A mechanism, causal explanation, or evaluative score \\
Motif / symbol & Recurrent textual or imagistic feature & A standalone interpretation of its significance \\
Skeleton & Curated set of work--axiom mappings & The work's unique semantic representation \\
\bottomrule
\end{tabularx}
\end{table}

\subsection{Artifact structure}

The analyzed snapshot consists of \texttt{axioms\_consolidated.json} and \texttt{relationships.json}. Axiom statements, descriptions, domains, keywords, examples, work titles, mapping notes, and relationship notes are primarily in Chinese. English examples in this paper are author translations for exposition.

The consolidated file has two arrays. The first stores axiom records; the second stores mappings $(work, axiom, role, note)$. Mapping roles are \emph{primary}, \emph{secondary}, and \emph{technique}. The relationship file contains typed inter-axiom links. Full literary source texts are not part of these artifacts.

\subsection{Typed relationships}

The relationship file stores a link as $(a_i,a_j,r,w,n)$, with source and target identifiers, a relation label $r$, a numeric strength $w$, and a note $n$. The five labels are:

\begin{description}[leftmargin=8.5em,style=nextline]
  \item[\normalfont\scshape Complements] the two claims illuminate compatible aspects of a concern;
  \item[\normalfont\scshape Tensions] the claims pull toward competing interpretations without being treated as strict negations;
  \item[\normalfont\scshape Specializes] one endpoint narrows the scope of the other;
  \item[\normalfont\scshape Evolves\_from] the note proposes a historical or conceptual development;
  \item[\normalfont\scshape Contradicts] the claims are curated as difficult to hold together in the same interpretive frame.
\end{description}

Complement, tension, and contradiction are conceptually symmetric even though the JSON stores ordered endpoints. Specialization and evolution require direction. The current strength values are curator-supplied confidence-like annotations, not probabilities or empirical scores; their mean of 0.839 should therefore be read descriptively only.

For example, the dataset links \axiom{036}, translated as ``Excess cognition can paralyze action,'' with \axiom{163}, ``A person may be destroyed without being defeated,'' using \textsc{Tensions}. The link records a possible contrast between reflective hesitation and resilient action. It does not assert that either claim is universally true.

\paragraph{Domain.} Eleven primary domains organize the collection. The data contain 74 observed domain/sub-domain pairs but 71 distinct sub-domain strings because several sub-domain labels occur under more than one primary domain. Table~\ref{tab:domains} gives the verified primary-domain distribution; English labels are translations of the Chinese metadata.

\begin{table}[t]
\centering
\caption{Primary-domain distribution in the dataset snapshot.}
\label{tab:domains}
\small
\begin{tabular}{@{}lrr@{}}
\toprule
\textbf{Domain} & \textbf{Count} & \textbf{Share} \\
\midrule
Human nature & 246 & 16.9\% \\
Society & 180 & 12.4\% \\
Existence & 172 & 11.8\% \\
Narrative structure & 156 & 10.7\% \\
Morality & 128 & 8.8\% \\
Emotion & 119 & 8.2\% \\
Language features & 119 & 8.2\% \\
Relationships & 88 & 6.0\% \\
Perspective & 85 & 5.8\% \\
Emotional tone & 82 & 5.6\% \\
Style and school & 80 & 5.5\% \\
\midrule
Total & 1,455 & 100.0\% \\
\bottomrule
\end{tabular}
\end{table}

\subsection{Data dictionary and unit of analysis}
\label{sec:data-dictionary}

The unit of analysis is an \emph{axiom record}, not a sentence in a source text. The record is designed to preserve enough structure for retrieval while keeping interpretive justification visible. Table~\ref{tab:data-dictionary} lists the record fields and associated collections relevant to analysis. Field names follow the JSON artifact; the final row names the separate relationship file's array, not a field in the consolidated file. Translations are explanatory labels used in this paper.

\begin{table}[t]
\centering
\caption{Data dictionary for axiom records and associated collections.}
\label{tab:data-dictionary}
\scriptsize
\begin{tabularx}{\linewidth}{@{}l l Y@{}}
\toprule
\textbf{Field} & \textbf{Type} & \textbf{Interpretation and audit expectation} \\
\midrule
\texttt{ax\_id} & string & Stable identifier; unique within the snapshot. \\
\texttt{statement} & string & Concise proposition-like formulation; not a truth value. \\
\texttt{description} & string & Qualification, rationale, or elaboration of the statement. \\
\texttt{category} & enum & Content or form; mixed cases require an explicit curation decision. \\
\texttt{type} & enum & Intended contextual scope (A--D); current C/D population is unresolved. \\
\texttt{abstraction} & enum & Fundamental, intermediate, or specific indexing level. \\
\texttt{domain}, \texttt{sub\_domain} & strings & Controlled vocabulary labels, currently translated for reporting. \\
\texttt{era\_context}, \texttt{region\_context} & lists & Declared scope qualifiers; empty lists do not prove universality. \\
\texttt{keywords} & list[string] & Retrieval aids, not a semantic decomposition. \\
\texttt{literary\_examples} & list[object] & Work title plus a curator's \texttt{how} grounding note. \\
\texttt{mappings} & list[object] & Top-level array in the consolidated file: \texttt{work\_title}, \texttt{ax\_id}, \texttt{role}, and \texttt{deduction\_note}. \\
\texttt{relationships.json} & list[object] & Separate file: \texttt{from}, \texttt{to}, \texttt{relation\_type}, \texttt{strength}, and \texttt{note}; strength is curator supplied. \\
\bottomrule
\end{tabularx}
\end{table}

This choice of unit has consequences. A single work may be represented by several records at different abstraction levels, while one record may in principle be mapped to several works. The current snapshot is highly work-local (Section~\ref{sec:audit}), so it should not be treated as a balanced ontology of literary concepts. The data dictionary is also intentionally permissive about natural language: enforcing a formal grammar would make automated validation easier but would exclude the ambiguity and qualification that human readers use to distinguish a useful claim from a slogan.

\subsection{Construction claims and provenance limits}
\label{sec:provenance}

Project documentation describes an iterative process in which language models proposed candidate formulations and human curators selected, rewrote, classified, deduplicated, and grounded records. The editorial criteria were single-focus formulation, non-triviality, usefulness for interpretation or creation, and connection to a named literary example.

The repository snapshot, however, does not preserve the evidence needed to reconstruct this pipeline exactly. It lacks a versioned candidate pool, model and prompt manifests for every batch, source-passage identifiers, curator identities and decisions, and adjudication logs. We therefore do not report candidate counts, time spent, number of curators, rejection percentages, or model-specific behavior. The present work is a description and audit of the resulting artifact, not a reproducible extraction experiment. Future releases should preserve immutable source records and complete provenance for every transformation.

\subsection{A provenance-aware annotation procedure}
\label{sec:annotation-procedure}

The following procedure is a proposed protocol for future releases, reconstructed from the needs of the schema rather than presented as a record of how the current snapshot was produced. It separates interpretation from normalization and makes disagreement recoverable.

\begin{enumerate}[leftmargin=2em]
  \item \textbf{Declare the reading context.} Record the work edition or translation, language, annotator identity or pseudonym, date, and the interpretive question. A claim without a declared context should be marked as context-unknown rather than silently assigned Type A.
  \item \textbf{Draft a candidate claim.} Write one proposition in plain language, then add a qualification explaining whose perspective, what textual pattern, or what formal choice makes the claim available. Avoid converting a plot summary or a moral verdict directly into an axiom.
  \item \textbf{Attach evidence and alternatives.} Add one or more source-passage identifiers where licensing permits, a short evidence note, and at least one plausible alternative reading. The alternative is not required to be stored as another axiom; its purpose is to prevent the grounding note from presenting a contestable inference as an observation.
  \item \textbf{Normalize and classify.} Assign category, context scope, abstraction, domain, sub-domain, and keywords. A second annotator should independently classify a sample before labels are merged. Mixed content/form claims may receive one primary category plus a note rather than being duplicated.
  \item \textbf{Map with a role.} Link the record to a work as primary, secondary, or technique, and explain why that role is appropriate. Negative mappings and rejected candidates should be retained in an adjudication log even if they are absent from the release artifact.
  \item \textbf{Adjudicate and version.} Preserve both original and edited wording, the reason for each edit, and the final decision. Relationship strengths should be accompanied by a rubric or left missing; a numeric value without a decision rule should not be interpreted as calibrated confidence.
\end{enumerate}

This protocol is intentionally more demanding than the current artifact. It would support audit trails, multilingual review, and held-out evaluation, but it also increases annotation cost. The central design choice is to preserve provenance as a first-class object rather than attempting to make the final statement appear objective by removing its history.

\subsection{Descriptive audit}
\label{sec:audit}

We recomputed every number in this section directly from the JSON files. Standard-library Python scripts stored with the manuscript run the integrity audit and generate Figure~\ref{fig:distributions}. Table~\ref{tab:summary} reports the snapshot statistics.

\begin{table}[t]
\centering
\caption{Verified descriptive statistics. Percentages may not sum exactly because of rounding.}
\label{tab:summary}
\small
\begin{tabular}{@{}lr@{}}
\toprule
\textbf{Statistic} & \textbf{Value} \\
\midrule
Axiom records & 1,455 \\
Content / form & 933 (64.1\%) / 522 (35.9\%) \\
Type A / B / C / D & 1,189 / 96 / 47 / 123 \\
Fundamental / intermediate / specific & 482 / 871 / 102 \\
Primary domains / distinct sub-domain labels & 11 / 71 \\
Literary-example records & 1,464 \\
Work--axiom mappings & 1,464 \\
Unique mapped works & 149 \\
Primary / secondary / technique mappings & 432 / 507 / 525 \\
Inter-axiom links & 472 \\
Complement / tension / specialization & 350 / 55 / 39 \\
Evolution / contradiction & 21 / 7 \\
Mean link strength (range) & 0.839 (0.70--0.93) \\
\bottomrule
\end{tabular}
\end{table}

\input{dataset_distribution.tex}

The audit found 1,455 unique axiom identifiers. Every axiom has at least one literary example and one work mapping. All 1,464 mapping endpoints and all 944 relationship endpoints resolve to an axiom identifier. No duplicate $(work,axiom)$ pair or self-link occurs. Works have 8--13 mappings (median 10; mean 9.83). Only seven axioms map to more than one work, with a maximum of four mappings, so the present artifact is better characterized as a set of work-derived records than as a densely reused cross-work ontology.

The audit also identifies a taxonomy issue. All 1,189 Type A records have empty era and region lists. Of 96 Type B records, 95 have an era but no region and one has neither. All 47 Type C records and all 123 Type D records have both era and region lists populated. Consequently, the stored context fields do not operationally distinguish Types C and D, despite their intended definitions. This should be corrected or the type scheme revised before type labels are used in comparative analysis.

These checks establish structural consistency, not interpretive quality. They cannot determine whether a claim is insightful, whether its example is persuasive, or whether a work mapping would be accepted by other readers.

\section{Axiom Skeletons as Interpretive Representations}
\label{sec:mapping}

For a work $w$, define its \emph{axiom skeleton} under a particular annotation process as
\[
S_w=\{(a,r,n):(w,a,r,n)\text{ occurs in the mapping file}\}.
\]
The qualifier is essential: $S_w$ represents a curated reading, not the work's unique or complete meaning. Multiple annotators could produce overlapping but different skeletons.

Table~\ref{tab:gatsby} presents all 12 mappings stored for \emph{The Great Gatsby}. The English glosses closely translate the corresponding Chinese axiom statements; the role labels come directly from the mapping file.

\begin{table*}[t]
\centering
\caption{The stored axiom skeleton for \emph{The Great Gatsby}.}
\label{tab:gatsby}
\small
\begin{tabularx}{\textwidth}{@{}llY@{}}
\toprule
\textbf{ID} & \textbf{Role} & \textbf{Condensed English gloss} \\
\midrule
\axiom{001} & Primary & Desire ultimately pursues the temporal illusion carried by its apparent object. \\
\axiom{003} & Primary & A morally lucid observer may remain unable to prevent tragedy and become implicated in it. \\
\axiom{004} & Primary & When love is assigned a redemptive function, it can become a source of destruction. \\
\axiom{002} & Secondary & Accumulated wealth can reproduce class barriers through less visible cultural forms. \\
\axiom{005} & Secondary & Spiritual desolation can conceal itself as hedonism beneath material prosperity. \\
\axiom{006} & Secondary & Social ascent can require an identity reconstruction that exiles the earlier self. \\
\axiom{007} & Secondary & Refusal of compromise can make an idealist's fall more complete. \\
\axiom{008} & Secondary & Privileged indifference can be produced by class position rather than individual malice alone. \\
\axiom{009} & Technique & Limited first-person narration can create suspense and irony through restricted knowledge. \\
\axiom{010} & Technique & Recurrent symbols can form a lyrical logic beyond plot. \\
\axiom{011} & Technique & Satire can emerge from a controlled gap between narrative tone and narrated content. \\
\axiom{012} & Technique & Retrospective framing can create tragic inevitability when the outcome is already known. \\
\bottomrule
\end{tabularx}
\end{table*}

\subsection{Contrasting skeletons and what they preserve}
\label{sec:contrasting-skeletons}

The skeleton abstraction becomes more informative when works are compared without assuming that the claims are interchangeable. Table~\ref{tab:contrasting-skeletons} gives two compact slices of the stored mappings. The statements are translated from the records and shortened only for layout; the identifiers and role labels are unchanged. \emph{Hamlet} concentrates on epistemic hesitation, inherited obligation, and contamination through power, whereas \emph{To Live} concentrates on endurance, historical dispossession, and the contingency of death. The contrast is an observation about this curation, not a claim that either work has one definitive philosophical center.

\begin{table*}[t]
\centering
\caption{Contrasting slices of two stored axiom skeletons.}
\label{tab:contrasting-skeletons}
\small
\begin{tabularx}{\textwidth}{@{}llY llY@{}}
\toprule
\multicolumn{3}{c}{\textbf{\emph{Hamlet}}} & \multicolumn{3}{c}{\textbf{\emph{To Live}}} \\
\textbf{ID} & \textbf{Role} & \textbf{Condensed gloss} & \textbf{ID} & \textbf{Role} & \textbf{Condensed gloss} \\
\midrule
\axiom{036} & Primary & Excess cognition can paralyze action. & \axiom{058} & Primary & Living itself can remain the last ground of meaning. \\
\axiom{037} & Primary & The execution of revenge contaminates its justice. & \axiom{059} & Primary & Suffering need not purify or educate; it can simply occur. \\
\axiom{038} & Primary & Usurped power poisons the wider relational ecology. & \axiom{060} & Secondary & Historical forces can make individual fate appear negligible. \\
\axiom{040} & Secondary & A performed mask can blur into the face it conceals. & \axiom{061} & Primary & Kinship can give endurance a reason without preventing loss. \\
\axiom{042} & Secondary & If life is performance, behavior cannot certify truth. & \axiom{063} & Secondary & The contingency of death resists solemn human framing. \\
\axiom{045} & Technique & A play within a play makes spectators inspect their own looking. & \axiom{066} & Technique & A restrained, minimal voice lets cruel facts carry affect. \\
\bottomrule
\end{tabularx}
\end{table*}

The comparison illustrates three uses of explicit roles. Primary mappings identify claims that the stored reading treats as central; secondary mappings record supporting or adjacent claims; technique mappings preserve formal means without collapsing them into themes. It also illustrates why cross-work reuse must be evaluated, rather than inferred from similar English glosses. For example, both works contain loss, but the records place loss within different proposed mechanisms: epistemic and political contamination in \emph{Hamlet}, and historical contingency and endurance in \emph{To Live}. A future remapping study could test whether independent annotators retain this distinction.

The examples also show where the metaphor fails if taken too literally. The skeletons do not provide a proof that action is impossible, that suffering lacks meaning, or that kinship is universally sustaining. They offer handles for asking how a work stages these possibilities, which voices contest them, and which formal choices make the reading persuasive. Their proper output is therefore a set of inspectable questions, not a theorem about the work.

The representation supports transparent comparison: readers can dispute an individual formulation, role, or grounding note rather than only accepting or rejecting a holistic summary. It may also support retrieval by domain or technique. At present, however, cross-work claims should be made cautiously. Because 1,448 of 1,455 axioms map to exactly one work, co-occurrence analysis is sparse and largely reflects how records were created. A convincing demonstration of shared claims across traditions requires a remapping study in which annotators actively consider existing axioms for every work.

\section{Testable Predictions and a Validation Program}
\label{sec:predictions}

The dataset instantiates the vocabulary and supports structural checks. The
following predictions concern the proposed mechanisms and evaluative dimensions.
They require new evidence. We distinguish measurement feasibility, reader studies,
and creative-process studies so that agreement about a record is not mistaken for
validation of the theory.

\subsection{Five predictions and their possible failure}

\paragraph{H1: Relational organization supports coherence (T1--T2).}
Given comparable passages with similar premise content, versions that make the
attribution and relation of conflicting commitments intelligible should help
readers explain character decisions with textual evidence and fewer unsupported
assumptions than versions containing unmotivated shifts.
A study could compare original and minimally altered passages, with independent
checks that edits preserve fluency and local event plausibility. Readers would
explain designated decisions without being shown the configuration. Judges blinded
to condition would score whether each explanation accounts for the decision using
textual evidence, under preregistered criteria permitting multiple readings. This
explanatory adequacy score is the primary outcome; coherence and liking are separate
secondary ratings. Manipulation checks of relational organization should use a
separate panel, so the outcome does not simply restate the definition of coherence.
If the manipulation changes perceived organization but not explanatory adequacy,
with sufficiently precise estimates, this counts against the proposed mechanism.
Mere dislike of an intentional contradiction would not by itself refute it.

\paragraph{H2: Scope conditions moderate contextual reach (T3--T4).}
Premises whose applicability depends on knowledge or institutions absent from a
reader's context should show a larger reduction in grounded recognition across
contexts than premises judged less dependent on those conditions. Independent
scope annotation should precede reception measurement. Multilingual reader groups
would report recognition, explain its textual basis, and rate resonance separately.
Analyses should consider translation, prior familiarity, and access to contextual
information. A precise null or reversed association after these checks would
challenge the proposed reach advantage. This is a conditional prediction about
reception, not an ordering of the A--D labels or of literary merit.

\paragraph{H3: Configuration differences inform originality judgments (T1--T4).}
Relative to a declared comparison corpus, changes in premise relations or formal
realization should explain some variation in expert distinctiveness judgments
beyond theme overlap alone. A study would sample works sharing broad concerns,
obtain independent premise and relation annotations, and compare a theme-only
model with one including configuration and realization features on held-out works.
Corpus provenance and earlier traditions must be recorded. No reliable incremental
explanatory value, or gains that disappear after controlling for length and genre,
would weaken this prediction. Neither identifier novelty nor set size is a valid
substitute for distinctiveness.

\paragraph{H4: Realization affects recoverability (T2--T3).}
When a writing plan identifies an organizing commitment and its intended role,
versions that sustain relevant textual consequences should make that commitment
more recoverable to readers than versions that state a topic but detach it from
events or form. Plans should be registered before drafts are evaluated. Readers
would freely reconstruct claims without seeing the plans; separate judges would
assess semantic alignment and textual grounding using preregistered criteria.
Explicitness, fluency, and length need matched or measured controls. If stronger
realization is independently confirmed but grounded recovery does not improve,
the proposed transmission mechanism is challenged. Recovery and aesthetic liking
remain separate outcomes.

\paragraph{H5: Organized openness supports grounded plurality (T2--T4).}
Texts that sustain multiple attributable perspectives should support more distinct,
evidence-backed reconstructions than matched texts that resolve the issue into one
explicit assertion. They should also yield stronger grounding than versions made
unclear by arbitrary omissions. Multiple readings from individual readers and
readings across readers should be analyzed separately. Judges blinded to condition
would cluster semantically similar claims and rate the evidence, reporting diversity
and grounding separately. If apparent richness consists only of unsupported or
incoherent claims, the predicted advantage fails. With successful manipulation
checks and sufficiently precise estimates, no increase in grounded plurality over
the resolved version, or no grounding advantage over the arbitrary-omission
version, would also count against the corresponding prediction. This comparison
distinguishes productive openness from information loss.

\subsection{Stage 1: Expert validation of records}
\label{sec:evaluation}

No expert ratings or human-subject results are available in the repository. We therefore specify a protocol that can be executed in a subsequent study.

\paragraph{Sampling and raters.}

The existing evaluation design calls for five raters with complementary experience in literary scholarship, creative writing, and digital humanities. A 35-record stratified sample can serve as a feasibility study, deliberately oversampling Types B--D and balancing content/form, abstraction levels, and domains. Because this sample and panel are small, resulting estimates should be reported with uncertainty and treated as formative. A later confirmatory study should use a larger multilingual panel and a larger independently drawn sample.

Raters should receive axiom statements, descriptions, and examples in a language they can evaluate. For classification tasks, the original labels should be hidden. The protocol, sampling seed, translations, consent materials, de-identified ratings, and analysis code should be archived before results are added to the paper.

\paragraph{Tasks and analysis.}

Each sampled record can be rated on five-point ordinal scales for (i) single focus, (ii) clarity, (iii) grounding, (iv) context adequacy, and (v) analytic usefulness, using the rubric in Appendix~\ref{app:rubric}. Analytic usefulness includes whether the record adds a non-trivial question or decision beyond a broad theme. Raters should also identify whether the statement overgeneralizes from its example and provide a short alternative reading.

For taxonomy evaluation, raters independently assign category, context type, abstraction, and domain. For mapping evaluation, they judge the mapped role and whether the note provides sufficient evidence for the connection. Relationship evaluation should sample each relation type and ask whether the label, direction where applicable, and explanatory note are appropriate.

Ordinal Krippendorff's $\alpha$ can summarize multi-rater reliability for scaled judgments, while nominal $\alpha$ can be used for categorical assignments. Reports should include bootstrap confidence intervals, per-class confusion matrices, missing-data handling, and the full score distributions rather than only means. Interpretive disagreement is substantive evidence: low agreement may reveal underspecified labels or legitimate plural readings rather than annotator failure.

\paragraph{Decision criteria.}

The study should define revision rules before examining aggregate results. For example, records with weak grounding or recurrent overgeneralization can be rewritten or removed; confused category pairs can trigger taxonomy revision; and region-specific entries should be reviewed by scholars familiar with the relevant language and tradition. We intentionally set no numerical acceptance threshold here because none has been justified by pilot data.

\subsection{Stage 2: Reader and comparison studies}

After the instruments can distinguish scope, relation, realization, and grounding,
H1, H2, H4, and H5 can be studied with readers and H3 with a declared comparative
corpus. Preregistration should specify one primary outcome for each confirmatory
test, contrast definitions, sampling, exclusions, and analysis. Passages and readers
are both sources of variation: repeated judgments should be analyzed with models
that account for both, rather than treating every rating as an independent sample.
Sample sizes require pilot variance estimates and power or precision targets.
Report effect sizes and uncertainty, adjust for multiple confirmatory comparisons,
and preserve unsuccessful predictions and competing readings.

\subsection{Stage 3: Creative-process studies}

A separate intervention can compare configuration-guided planning with theme-based
planning under matched time, writing tools, and task constraints. Random assignment
or a counterbalanced design should account for writer experience and practice
effects. Plans and successive drafts would document selection, relational revision,
and realization; judges blinded to condition would assess completed texts. Evidence
for improved planning visibility is distinct from evidence for improved literary
quality. A useful outcome could be clearer revision decisions without higher
overall ratings; a null or adverse effect on the preregistered outcomes must remain
reportable. This study tests the framework as a creative method, beyond its use as
an interpretive representation.

\section{Relation to Existing Approaches}
\label{sec:related}

\subsection{Narrative structure and computation}

Structuralist narratology supplied influential vocabularies for recurring forms. Propp identified 31 functions in a corpus of Russian folktales \citep{propp1968morphology}; Greimas modeled relations among actants \citep{greimas1966semantique}; and Todorov described narrative transformation in grammatical terms \citep{todorov1969grammaire}. Lehnert linked affective states and events through plot units for narrative summarization \citep{lehnert1981plot}. Later hand-built story systems represented goals, plans, and reusable narrative cases \citep{meehan1977talespin,turner1994creative,perez2001mexica}. Statistical approaches learn event chains \citep{chambers2008unsupervised,pichotta2016learning}, while neural generators model longer-form story continuation \citep{fan2018hierarchical,see2019massively}. These traditions supply representations of action, transformation, and generative practice. The proposed contribution here is to connect conceptual premises and formal choices in a configuration that can also be reconstructed by readers and evaluated through the five dimensions. It builds on structured narrative representation rather than claiming that premise-guided creation is unprecedented.

\subsection{Computational literary studies and cultural data}

Distant-reading research demonstrates how structured evidence can support comparison across large literary collections \citep{moretti2005graphs,jockers2013macroanalysis,underwood2019distant}. Other work models characters and relationships \citep{bamman2014bayesian,chaturvedi2016modeling}, sentiment trajectories \citep{reagan2016emotional}, or emotion dynamics \citep{vishnubhotla2024emotion}. Cultural-heritage ontologies such as CIDOC CRM and FRBRoo provide rigorous descriptions of objects, events, and bibliographic entities \citep{doerr2003cidoc,bekiari2015frbr}. The implementation of \system stores curated interpretive assertions with examples and links. Its theoretical role is to make configurations inspectable across creation, interpretation, and evaluation; database structure alone does not establish that connection.

\subsection{Interpretation and paraphrase}

Reader-oriented accounts already locate meaning in the interaction between text
and reading. Iser's account of the reading process and Fish's account of
interpretive communities provide important antecedents for the situated
reconstruction proposed here \citep{iser1978act,fish1980text}. T3 does not claim
to discover reader plurality. It proposes an explicit unit for comparing some
of its outcomes: premises, relations, formal observations, and attributed scope.
This unit can make differences inspectable, while the originating reading
practices remain necessary to explain why those differences arise.

Encoding interpretation as propositions risks detaching meaning from form. Brooks's critique of the ``heresy of paraphrase'' remains directly relevant: a paraphrase cannot substitute for the organization and language of a literary work \citep{brooks1947well}. Our records are therefore indexes into possible readings, not compressed replacements for texts. The conceptual model makes form constitutive through $F$ and attribution through $\kappa$. Its synthesis joins this dependence on form to explicit premise selection, situated reconstruction, and comparative evaluation. The test is whether these connections illuminate decisions and disagreements beyond a thematic paraphrase; the proposed theme baseline tests incremental usefulness, not priority over all literary theory.

\section{Discussion}
\label{sec:discussion}

\subsection{What the formalism contributes}

The framework makes interpretive commitments inspectable. A conventional thematic tag can conceal whether two readers mean the same thing by ``alienation''; proposition-like records state a claim that can be compared, challenged, translated, linked, and applied. Typed mappings also separate claims treated as central to a reading from formal techniques used by a work.

Formalization may be useful in pedagogy, where students can compare skeletons and explain disagreements; in digital scholarship, where claims can be queried across works; and in computational creativity, where a writer can select or tension claims before developing characters and scenes. These are prospective applications. The present descriptive audit does not demonstrate learning gains, improved interpretation, or better generated literature.

\subsection{Cultural and epistemic limitations}

The artifact reflects canon selection, the languages and traditions available to its curators and language models, and a preference for explicit propositions. These choices may underrepresent oral, performative, indigenous, popular, and formally resistant traditions. Translation can erase ambiguity or import categories absent from a source tradition. A label such as context-independent may reflect curator distance from the context rather than actual universality.

The dataset also records interpretations without attribution to particular critics or reading communities. This makes claims convenient to reuse but weakens intellectual provenance and can make contested judgments appear neutral. Future versions should distinguish curator synthesis, cited scholarship, textual evidence, and model suggestion, with versioned attribution for each record.

\subsection{Technical and data limitations}

The current snapshot is structurally consistent but not a mature benchmark. It has no held-out annotation set, external validity results, calibrated relationship strengths, or documented negative examples. Sparse reuse across works limits graph and co-occurrence analysis. Type C and D are not distinguishable by populated context fields. The lack of complete prompts, model versions, source passages, and editorial histories prevents exact reconstruction of the dataset.

These limitations set the next research priorities: repair the context taxonomy; add work-edition and source-passage identifiers; preserve generation and adjudication provenance; remap works against a shared candidate pool; and complete multilingual expert validation before making claims about generality or quality.

\section{Ethics and Artifact Availability}

No completed expert study or user study is reported, and this descriptive work did not collect new participant data. A future expert study will require appropriate institutional review or exemption determination, informed consent, secure handling of identifiable responses, and explicit permission for release of de-identified ratings.

The dataset contains interpretive propositions, metadata, and explanatory examples rather than full copies of literary works. Nevertheless, future public release should review quotations, translations, and metadata for copyright and licensing requirements. Model-assisted formulations may reproduce representational biases or unattributed critical language; human review and provenance tracking are necessary before downstream deployment.

The two JSON files used for the reported audit are maintained in the authors' development repository. At the time of writing, no public archival identifier or explicit dataset license has been assigned. The audit and figure-generation scripts are stored with the manuscript, and all aggregate claims in this paper can be recomputed from the named JSON files within that repository. A versioned archive, license, data statement, and checksums should accompany any archival dataset release.

\section{Conclusion}
\label{sec:conclusion}

The literary-axiom framework connects creation, interpretation, and evaluation
through the organization and realization of conceptual and formal commitments.
Its central proposal is that a selected configuration becomes available through
literary choices and is reconstructed differently by situated readers. The model
connects three possibilities: shared premises can yield different works, coherent
literature can stage conflict, and multiple readings can combine openness with evidence.

The five evaluative dimensions articulate this proposal as a research program.
Coherence concerns organized relations, reach concerns transfer across contexts,
distinctiveness concerns comparison, fidelity concerns realization, and richness
concerns grounded plurality. Each has a proposed empirical test and a stated way
in which the prediction could fail. Together they preserve the original ambition
to reason systematically about literary creation while allowing form, conflict,
and cultural specificity to carry meaning.

The implementation supplies 1,455 records, mappings to 149 works, and typed
relationships as an inspectable starting point. Its audit establishes structural
properties; studies of readers, comparative judgments, and creative practice must
determine the framework's explanatory and practical value. The next step is to
test whether explicit configurations illuminate decisions and differences that
would otherwise remain hidden in broad thematic descriptions.

\section*{Author Contributions}

Qiang Liu: conceptualization, methodology, data curation, formal analysis, and
writing--original draft. Chunyi Zhao: domain-informed critical review from a
literary-studies perspective and writing--review and editing.

\clearpage
\appendix
\section{Proposed Annotation Rubric}
\label{app:rubric}

This appendix gives a compact instrument for the future validation study. It is a protocol artifact, not a result. Raters should score the record and its grounding note together, then write a brief reason for any score of 1 or 2. A ``not assessable'' response should be allowed when the rater cannot read the source language or the example lacks sufficient evidence.

\begin{table}[H]
\centering
\caption{Proposed five-point rubric for axiom records.}
\label{tab:rubric}
\scriptsize
\begin{tabularx}{\linewidth}{@{}lY Y@{}}
\toprule
\textbf{Criterion} & \textbf{High score (5)} & \textbf{Low score (1)} \\
\midrule
Single focus & One principal claim with scope that can be stated in one sentence. & Multiple unrelated claims or a slogan that cannot be scoped. \\
Clarity & A reader can identify the relation being proposed and what would count as a counter-reading. & Wording is metaphorical, circular, or too vague to disagree with. \\
Grounding & The note identifies concrete textual or formal features and explains the inferential step. & The note is a plot summary, unsupported verdict, or generic association. \\
Context adequacy & Historical, regional, linguistic, and edition limits are explicit where relevant. & A situated claim is presented as universal, or context is asserted without evidence. \\
Analytic usefulness & The record supports a question, comparison, or creative decision without replacing close reading. & The record adds no operation beyond a broad theme or forces a single interpretation. \\
\bottomrule
\end{tabularx}
\end{table}

For mapping review, raters can use the same scale for role fit and note adequacy, while separately selecting ``primary,'' ``secondary,'' ``technique,'' or ``not mapped.'' For relationship review, they should select the relation label, assess direction when applicable, and explain whether the note describes complementarity, tension, specialization, evolution, or contradiction. These tasks should be analyzed separately from record quality: a clear record can be mapped to the wrong work, and a plausible mapping can expose a poorly formulated record.

\end{document}

%% file: dataset_distribution.tex
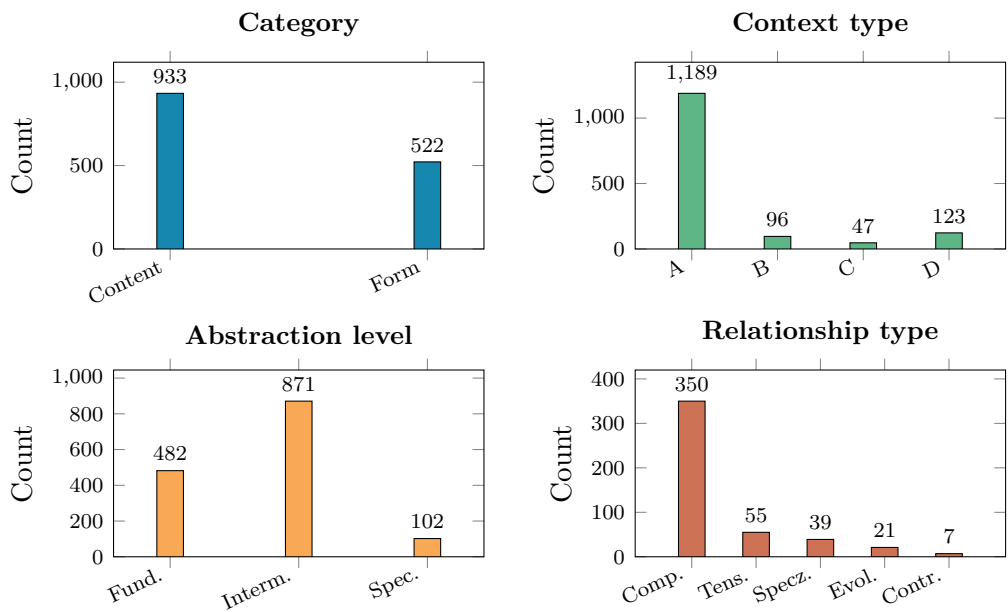
\begin{figure*}[t]
\centering
\begin{tikzpicture}
\begin{groupplot}[
  group style={group size=2 by 2, horizontal sep=2.0cm, vertical sep=1.6cm},
  width=0.40\textwidth,
  height=0.25\textwidth,
  ybar,
  ymin=0,
  enlarge y limits={upper,value=0.20},
  nodes near coords,
  nodes near coords style={font=\scriptsize},
  tick label style={font=\scriptsize},
  title style={font=\small\bfseries},
  enlarge x limits=0.22,
  ylabel={Count},
  symbolic x coords={Content,Form,A,B,C,D,Fund.,Interm.,Spec.,Comp.,Tens.,Specz.,Evol.,Contr.},
  xtick=data,
  x tick label style={rotate=25,anchor=east}
]
\nextgroupplot[title={Category}]
\addplot[fill=MidnightBlue!70] coordinates {(Content,933) (Form,522)};
\nextgroupplot[title={Context type}]
\addplot[fill=ForestGreen!65] coordinates {(A,1189) (B,96) (C,47) (D,123)};
\nextgroupplot[title={Abstraction level}]
\addplot[fill=BurntOrange!75] coordinates {(Fund.,482) (Interm.,871) (Spec.,102)};
\nextgroupplot[title={Relationship type}]
\addplot[fill=BrickRed!65] coordinates {(Comp.,350) (Tens.,55) (Specz.,39) (Evol.,21) (Contr.,7)};
\end{groupplot}
\end{tikzpicture}
\caption{Distributions recomputed from the checked-in JSON snapshot. Context types A--D are defined in Section~\ref{sec:taxonomy}; relationship labels are abbreviated for space.}
\label{fig:distributions}
\end{figure*}